\documentstyle[12pt]{article}
\begin{document}
\def\thefootnote{\fnsymbol{footnote}}
\begin{flushright}
KANAZAWA-05-09  \\
August, 2005\\  
\end{flushright}
\vspace*{2cm}
\begin{center}
{\LARGE\bf Dark world and baryon asymmetry from a common source}\\
\vspace{1 cm}
{\Large  Daijiro Suematsu}
\footnote[1]{e-mail:suematsu@hep.s.kanazawa-u.ac.jp}
\vspace {0.7cm}\\
{\it Institute for Theoretical Physics, Kanazawa University,\\
        Kanazawa 920-1192, JAPAN}
\end{center}
\vspace{2cm}
{\Large\bf Abstract}\\
We study generation of baryon number asymmetry and both abundance of 
dark matter and dark energy on the basis of global symmetry and its
associating flat directions in a supersymmetric model.
We assume the existence of a model independent axion which is generally 
expected in the effective theory of superstring.
If we consider a combined field of the model independent axion and
a pseudo Nambu-Goldstone boson coming from spontaneous breaking of
the global symmetry, its potential can be sufficiently flat and 
then it may present a candidate of the dark energy as a quintessential 
axion. Both the baryon asymmetry and the dark matter are supposed to be 
produced nonthermally as the asymmetry of another global charge 
through the Affleck-Dine mechanism along the relevant flat direction. 
Its decay to the observable and hidden sectors explains the baryon 
number asymmetry and the dark matter abundance, respectively. 
 
\newpage
\setcounter{footnote}{0}
\def\thefootnote{\arabic{footnote}}
\section{Introduction}
Recent observations of the anisotropy of cosmic microwave 
background \cite{cmb}, the distance-redshift relation of type Ia 
supernovae \cite{sna} and the large scale structure \cite{lss} show 
that there is a large amount of energy, called dark energy, 
which has negative pressure to cause acceleration of the expansion 
of the universe. 
Analyses of these data also allow us to determine energy density 
parameters with good accuracy. 
This situation for these astrophysical observations makes us confront
several difficult problems, that is, what is the origin of the dark 
energy, why each energy density parameter 
takes the present values and so on. 
In particular, since each energy density is expected to have different
origin and also behaves in a very different manner 
during the expansion of the universe, it seems to be very mysterious why 
they take similar order values in the present universe. 
The investigation of these problems might give us some clues to access 
new physics beyond the standard model of particle physics.
In the supersymmetric framework, especially,
many works have been done mainly related to the thermal neutralino dark 
matter \cite{bbd}.

As one of interesting approaches to these problems, one may consider
them on the basis of some global symmetry.
As proposed in \cite{dark}, both abundance 
of the baryon and the dark matter may be understood from a view point of
the charge asymmetry of a suitable common global symmetry. 
Although this idea seems to be very elegant and also be able to explain 
naturally why they have the similar order values in the present
universe, realistic models in this direction seem not to have been 
constructed a lot in the context of a supersymmetric framework.\footnote{An 
example of this kind of recent works may be found in \cite{hmw}.
We can also find several other attempts to explain both the dark matter 
and baryon asymmetry based on a single source. 
See, for example, \cite{single,nonth}.
We should note that the tuning of the dark matter mass to the proton mass
can remain as the unsolved problem even in the case that both number
density of the baryon and the dark matter are related each other through
a certain symmetry.}

The Lagrangian of the minimal supersymmetric standard model (MSSM) has only 
baryon number $(B)$ and lepton number $(L)$ as its global symmetry.
Even if we put aside the dark energy, it seems to be difficult to explain
both abundance of the baryon and the dark matter on the basis of 
these global symmetries. However, if we introduce new 
singlet chiral superfields in the MSSM, the global symmetry can be 
extended and then the idea given in \cite{dark} may be applicable to 
the supersymmetric model by using those symmetries. 
In the previous paper \cite{s}, we have discussed 
such a possibility to explain both abundance of the baryon and the dark matter 
on the basis of a common global symmetry. 
The global charge asymmetry induced by the
Affleck-Dine mechanism \cite{ad} in a $D$-flat direction associated with the
introduced singlet fields is liberated into the observable and hidden 
sectors. These decay products are shown to explain the baryon asymmetry 
and the dark matter abundance, 
respectively. Although we left the dark energy untouched in that study,
it seems to be interesting to investigate a new possibility 
that the dark energy 
may be explained in the same framework.   

For the dark energy, many works have been done by assuming the existence
of the slowly rolling homogeneous scalar field with extremely small
mass and a large vacuum expectation value, 
which is called quintessence \cite{quin,quin0,quin1}. 
Although they may be able to explain the observational data to some level,
many of them seem not to be motivated from particle physics.
Moreover, if we try to consider the quintessence in the supersymmetric
framework, it is feared that the supergravity corrections 
violate the slow roll nature of the quintessence as in the
case of inflaton \cite{sgra}. To escape this problem, 
we may impose some symmetry which can protect the flatness of 
the potential from such
corrections. As such a solution, one may consider that the quintessence
field is a pseudo Nambu-Goldstone (pNG) boson of some global 
symmetry \cite{aquin0}.
In that case, the flatness of its potential is guaranteed by the 
shift symmetry of the pNG field. Thus, the axion type field seems to 
be a promising candidate of the dark energy, which can be also supported by
a suitable motivation from particle physics. 
In such a scenario, a crucial problem is how we can realize the tiny mass
scale suitable for the dark energy. On this point, there is an interesting 
proposal in which a model independent axion in the superstring is used
as a candidate for the quintessence \cite{aquin}.

In this paper we try to extend the idea given in
\cite{aquin} so that the scenario for the baryon asymmetry and 
the dark matter abundance 
given in \cite{s} is embedded into the same framework.
If it is possible,  we may have a common source for the baryon asymmetry,
the dark matter and the dark energy. 
The remaining parts of this paper are organized as follows. 
In section 2 we present a supersymmetric model which is mainly 
characterized by the structure of the hidden sector. 
Several features of the model are discussed.
In section 3 the dark energy is studied in this model. We show that
two axion fields associated to the almost flat directions 
can have hierarchical mass and they can play the role of the quintessence
and the invisible axion for the strong CP problem, respectively.
In section 4 we discuss how both abundance of the baryon and the dark matter
can be produced in this framework. 
Section 5 is devoted to the summary. Some details of the calculation 
are explained in two Appendices.

\normalsize
\section{A model with $D$-flat directions}
We consider a supersymmetric model with observable and hidden sectors.
The observable sector is supposed to be composed of the MSSM contents
and the MSSM singlet chiral superfields
$\hat N_j$, $\hat S_1$. The superpotential is written as
\begin{equation}
W_{\rm ob}=\sum_{i,j=1}^3\left(h_u^{ij}\hat Q_i\hat{\bar U}_j\hat H_2
+h_d^{ij}\hat Q_i\hat{\bar D}_j\hat H_1
+h_e^{ij}\hat L_i\hat{\bar E}_j\hat H_1 
+h_\nu^{ij}\hat L_i\hat N_j\hat H_2\right)
+\lambda_1\hat S_1\hat H_1\hat H_2,
\end{equation}
which is similar to the usual next MSSM superpotential up to a cubic term
of $\hat S_1$.
The hidden sector is assumed to be composed of two parts.
One is a super Yang-Mills part composed of an SU(N) vector 
superfield and three sets of chiral superfields such as
$\hat F^\alpha~({\bf N}, \alpha=1,\cdots, n_{G_1}+n_{G_2})$, 
$\hat G_1^\beta~(\bar{\bf N}, \beta=1,\cdots, n_{G_1})$ 
and $\hat G_2^\gamma~(\bar{\bf N},
\gamma=1,\cdots, n_{G_2})$, where ${\bf N}$ and $\bar{\bf N}$ 
stand for the fundamental representation of SU(N) and its conjugate
representation, respectively. 
We assume that the supersymmetry is broken in this part of the hidden sector.
Another part is composed of several gauge singlet chiral superfields 
$\hat S_2$, $\hat C_i (i=1,\cdots,5)$, $\hat\phi$ and 
$\hat\psi$.\footnote{In this paper, we
put a hat on the character for the chiral superfields in both sector.
For the scalar component of the hidden
sector chiral superfield, we put a tilde on it. 
For its fermionic component, we use the same character 
as the superfield without the hat.} 
The superpotential for these chiral superfields is assumed to 
be\footnote{This part is relevant to the explanation of the dark matter
abundance as seen below. Although it may be possible to consider the 
similar structure to the observable sector for this part, 
we assume this simple structure here, for simplicity.} 
\begin{eqnarray}
&&W_{\rm hid}^{\rm YM}=\sum_{\beta=1}^{n_{G_1}}{g_1\over M_{\rm pl}}
\hat\phi^2 \hat F^\beta\hat G_1^\beta
+\sum_{\gamma=1}^{n_{G_2}}g_2\hat\psi\hat F^\gamma\hat G_2^\gamma, \nonumber\\
&&W_{\rm hid}^{\rm S}=\lambda_2\hat S_2 \hat C_1\hat C_2 
+{f_1\over M_{\rm pl}}\hat\phi^2 \hat C_1\hat C_3 
+{f_2\over M_{\rm pl}}\hat\phi^2 \hat C_2\hat C_4 
+{f_3\over M_{\rm pl}}\hat\phi^2 \hat S_2\hat C_5,
\end{eqnarray}
where $M_{\rm pl}$ is the reduced Planck mass and 
all coefficients of these operators are supposed
to be real and $O(1)$.  
The observable and hidden sectors are assumed to be connected only through the 
gravitational interactions which are represented by
the following superpotential:
\begin{equation}
W_{\rm mix}={d_1\over M_{\rm pl}^2}\hat S_1^2\hat S_2^3
+\sum_{\gamma=1}^{n_{G_2}}{d_2\over M_{\rm pl}}
\hat F^\gamma\hat G_2^\gamma\hat H_1\hat H_2,
\label{eqa}
\end{equation}
where the coefficients $d_{1,2}$ are also assumed to be real and $O(1)$. 
The operators in $W_{\rm mix}$ play essential roles in the present 
scenario as discussed below. 

Many singlet chiral superfields are introduced in the hidden sector.
This seems to be unavoidable to realize several physical results
simultaneously keeping the global symmetry in the hidden sector 
which is essential for the present approach. Although the detailed
discussions will be presented in the followings, it may be useful to
present a brief summary of their specific roles here.
The global charge asymmetry related to both the dark matter and baryon
asymmetry is generated through the flat direction of scalar components
of $\hat S_{1,2}$. $\hat\psi$ contributes to the supersymmetry breaking 
in the observable sector through its coupling with $\hat F^\alpha$ 
and $\hat G_2^\gamma$ in $W_{\rm hid}^{\rm YM}$. 
The vacuum expectation value of the scalar
component of $\hat\phi$ generates the mass of dark matter candidates 
$\hat C_i$ without violating the global symmetry through the
couplings in $W_{\rm hid}^{\rm S}$. The invariance of these couplings 
under the global symmetry requires to introduce 
a rather large number of $\hat C_i$.  
$\hat\phi$ also contributes to the generation of
the dark energy through the coupling in $W_{\rm hid}^{\rm YM}$.

In order to control the structure of $W_{\rm hid}$ and $W_{\rm mix}$, 
we may impose discrete symmetry $Z_{18}\times Z_3$. 
We assign their charges for each chiral superfield 
in the observable and hidden sectors as shown in Table 1 and 2. 
Each operator in $W_{\rm ob}$, $W_{\rm hid}$ and $W_{\rm mix}$
is $Z_{18}\times Z_3$ invariant. 
The operators in $W_{\rm hid}$ and $W_{\rm mix}$ are the lowest order 
ones which can be constructed in a
$Z_{18}\times Z_3$ invariant way by using the fields in both 
sectors.\footnote{Among gauge invariant 
renormalizable operators constructed from the contents given in Table 1 and 2, 
we cannot exclude $\hat S_1^3$ by imposing $Z_{18}\times Z_3$ only.
We need an additional symmetry like $Z_2$ to forbid it. 
Although its charge can be consistently assigned to all fields, we do
not present it explicitly here and only assume implicitly that such a 
symmetry is imposed.}
It can be checked that this $Z_{18}\times Z_3$ has no anomaly with 
respect to both the SM gauge group and SU(N) if we take $n_{G_1}=3$ 
and $n_{G_2}=1$. 

\begin{figure}[tb]
\begin{center}
\begin{tabular}{c|ccccccccccc}
 & $\hat{Q}_L$ &$\hat{\bar U}_L$ &$\hat{\bar D}_L$ & $\hat{L}_L$ 
 &$\hat{\bar E}_L$ &$\hat{H}_1$ &$\hat{H}_2$ & $\hat{S}_1$ &$\hat{N}$ &
 $A_3$ & $A_2$ \\ \hline
$Z_{18}$ & $1$ & $5$ & $5$ & $1$ & $5$ 
& $-6$ & $-6$ & $12$ & $5$ & 0 & 0 \\
$Q_{\rm PQ}$ &  0 &$-2$ &1 & $-1$ &2 &$-1$ & 2 &  $-1$ & $-1$ &
$-{3\over 2}$ & $-1$ \\
$Q_R$  & ${5\over 6}$ &$0$ & 1 &${1\over 6}$ 
&${5\over 3}$ &${1\over 6}$ & ${7\over 6}$ &${2\over 3}$ &${1\over 3}$ &
$-5$ & $-{13\over 3}$\\
$B$  & ${1\over 3}$  &$-{1\over 3}$ & $-{1\over 3}$ & 
0 & 0 & 0 &0 & 0 & 0 & 0 & ${3\over 2}$ \\
$L$  & 0   & 0   & 0  & $1$   & $-1$ & 0 & 0 & 0 & 0 &
0 & ${3\over 2}$ \\ 
$X$ & ${5\over 6}$ & ${59\over 9}$ & $-{22\over 9}$ & ${23\over 6}$ &
$-{16\over 3}$ & ${7\over 2}$ & $-{11\over 2}$ & 3 & ${11\over 3}$ & 0 & 0\\
$\tilde X$ & ${5\over 6}$ & ${26\over 3}$ & $-{10\over 3}$ & ${9\over 2}$ &
$-7$ &  ${9\over 2}$ & $-{15\over 2}$ & 5 & ${14\over 3}$ & ${3\over 2}$
 & 0 \\\hline
\end{tabular}
\vspace*{3mm}\\
\end{center}
{\footnotesize Table 1.~ A charge assignment for the chiral superfields in the
 observable sector and the anomaly $A_3$ and $A_2$ for SU(3) and SU(2).
A trivial $Z_3$ charge is assigned for all fields.} 
\end{figure}

\begin{figure}[tb]
\begin{center}
\begin{tabular}{c|cccccccccccc} 
 &$\hat F$ & $\hat G_1$& $\hat G_2$ & $\hat S_2$ & $\hat C_1$
& $\hat C_2$ & $\hat C_3$ & $\hat C_4$ & $\hat C_5$
& $\hat\phi$ &$\hat\psi$  & $A_N$  \\ \hline
$Z_{18}$ &0 & $-4$ & $-6$ &4 &$-3$ &$-1$ &$-1$ 
&$-3$ &$-8$ & 2 & 6 & $-{1\over 2}(4n_{G_1}+6n_{G_2})$  \\ 
$Z_3$ &2 & 0 &1 &0 &1 &2 &1 &0 &2 & 2 & 0& ${1\over 2}(2n_{G_1}+3n_{G_2})$  \\ 
$Q_{\rm PQ}$ & $-1$ & $1$ &  $0$ & ${2\over 3}$ & $-1$ 
& ${1\over 3}$ & 1 & $-{1\over 3}$ & $-{2\over 3}$ & 0 & 1& 
$-{1\over 2}n_{G_1}$   \\
$Q_R$ & ${2\over 3}$ & ${4\over 3}$ & 0 & $-{10\over 3}$ 
& ${2\over 3}$ & ${14\over 3}$ & ${4\over 3}$ & $-{8\over 3}$ 
& ${16\over 3}$ &0 & ${4\over 3}$ & $-{2\over 3}n_{G_2}-N$  \\
$X$ & $4$ & $-2$ &0 & $-{10\over 3}$  & 4 & ${4\over 3}$ 
&$-2$ & ${2\over 3}$ & ${16\over 3}$ & 0 & $-2$ & $n_{G_2}-N$ \\
$Y$ & 12 & $-10$ & 0 & $-{46\over 3}$ & 12 & ${16\over 3}$ & $-10$ &
 $-{10\over 3}$ & ${52\over 3}$ & 0 & $-10$ & $5n_{G_2}-N$ \\ 
$\tilde X$ & 5 & $-3$ & 0 & $-4$  & 5 & ${29\over 9}$ &
$-3$ & $-{11\over 9}$ & 6 & 0 & $-3$ & ${3\over 2}n_{G_2}-N$  \\\hline
\end{tabular}
\vspace*{3mm}\\
\end{center}
{\footnotesize Table 2.~ A charge assignment for the chiral superfields in the
hidden sector and the anomaly $A_N$ for SU(N). All fields have $B=L=0$.} 
\end{figure}

In the MSSM, both operators $\hat{H}_1\hat{H}_2$ and $\hat{L}_i\hat{H}_2$ are 
gauge invariant. Thus, as shown in $W_{\rm ob}$, we can construct two 
gauge invariant dimension three operators $\hat{S}_1\hat{H}_1\hat{H}_2$ 
and $\hat{L}_i\hat{N}_j\hat{H}_2$ by introducing the singlet chiral 
superfields $\hat{S}_1$ and $\hat{N}_j$.
If we add $\hat{S}_1\hat{H}_1\hat{H}_2$ to the MSSM superpotential 
and remove the ordinary $\mu$-term as in $W_{\rm ob}$, 
the global symmetry is generally extended.
In fact, in the present model two new global Abelian symmetries appear
in addition to the $B$ and $L$ symmetries as far as we restrict our target 
to the renormalizable operators \cite{iq,dr,s}.  
They are taken as the Peccei-Quinn type symmetry and the $R$
symmetry.\footnote{Since the ordinary Peccei-Quinn symmetry has no SU(2)
gauge anomaly, the present one is not such a symmetry in an exact sense.}
We present an example of their charge assignment for each chiral
superfield in the observable and hidden sectors in Table 1 and 2, 
respectively. Gauginos are considered to have $Q_R=-1$ only.
We note that the global symmetry in the hidden and observable sectors 
can be related each other since the model is assumed to have 
the operators in $W_{\rm mix}$ .
This charge assignment will be used in the following discussion.

Among these four global Abelian symmetries, 
U(1)$_{B-L}$ and U(1)$_X$ remain as those with no SU(3) and SU(2) 
gauge anomalies in the observable sector. 
The U(1)$_X$ symmetry is a kind of $R$-symmetry and
its charge can be taken as the linear combination of the four global 
Abelian charges in such a way as $X={1\over 3}(B+L-10Q_{\rm PQ})+Q_R$.
In the hidden sector, there is also an SU(N) anomaly free global Abelian
symmetry in the case of $n_{G_1}=3$, $n_{G_2}=1$ and $N=5$.
This charge can be represented as $Y=X-8Q_{\rm PQ}$. 
These charges are listed in Table 1 and 2.
It may be interesting to note that the same values of $n_{G_1}$ and $n_{G_2}$ 
make $Z_{18}\times Z_3$ be gauge anomaly free as mentioned before.

It is useful to note that only the soft supersymmetry breaking operator 
associated with the first term in $W_{\rm mix}$ can violate 
the $X$ conservation in the Lagrangian derived from 
$W_{\rm ob}$, $W_{\rm hid}$ and $W_{\rm mix}$. 
As we will see later, it can play an essential role to generate the $X$
asymmetry through the AD mechanism in the $D$-flat direction of the
scalar components of $\hat S_{1,2}$. 

Using U(1)$_{\rm PQ}$ and U(1)$_R$, we can also construct a global 
Abelian symmetry U(1)$_{\tilde X}$, which is SU(2) anomaly free but have
SU(3) and SU(N) anomalies. As such an example, we can consider 
$\tilde X=-{13\over 3}Q_{\rm PQ} + Q_R$. In the next section, 
we will suggest that this global symmetry
can be related to the appearance of a quintessential axion. 

Here we make some assumptions for the dynamics in the super Yang-Mills
part of the hidden sector along the line of 
the arguments given in \cite{aquin}.
The SU(N) gaugino $\lambda_h$ and the scalar components $\tilde F$, 
$\tilde G_{1,2}$ of the chiral superfields $\hat F$, $\hat G_{1,2}$ 
are supposed to condense at the scale $\Lambda_h$. 
The condensations of the gaugino, the scalar components 
$\tilde F$ and $\tilde G_{1,2}$ are defined by 
\begin{equation}
\langle\lambda_h\lambda_h\rangle\equiv
\Lambda_h^3 e^{i{\theta_\lambda(x) \over \Lambda_h}}, \qquad
\langle\tilde F^\beta\tilde G_1^\beta\rangle\equiv
\Lambda_h^2 e^{i{\theta_{G_1}(x) \over \Lambda_h}}, \qquad
\langle\tilde F^\gamma\tilde G_2^\gamma\rangle\equiv
\Lambda_h^2 e^{i{\theta_{G_2}(x) \over \Lambda_h}},
\label{cond}
\end{equation}
where $\beta=1,\cdots,n_{G_1}$ and $\gamma=1,\cdots,n_{G_2}$.
Although these condensations do not break the the supersymmetry by themselves, 
the U(1)$_X$ and U(1)$_{\tilde X}$ symmetries are spontaneously 
broken in the hidden sector by these condensations\footnote{It should
be noted that these breakings have no substantial effect to the singlet
part of the hidden sector, which will be discussed later.} and 
the NG bosons $\theta_{G_{1,2}}(x)$ and $\theta_\lambda(x)$ appear 
as composite states.
Since U(1)$_{\tilde X}$ has SU(3) and SU(N) anomalies, 
the corresponding NG boson gets a mass through these effects to be the 
pNG boson.\footnote{  
The scenario for the dark energy in \cite{aquin} is regarded to be
constructed from this pNG and the model independent axion 
$a_{\rm MI}(x)$.} 
It can be expressed by using $\theta_{G_1},~\theta_{G_2}$ and $\theta_\lambda$ 
as \cite{cp,scp} 
\begin{equation}
\theta_{\tilde X}= {2n_{G_1}\theta_{G_1}+5n_{G_2}\theta_{G_2}
-2\theta_\lambda \over (4n_{G_1}^2+25n_{G_2}^2+4)^{1/2}}.
\end{equation}

The supersymmetry breaking in the observable sector 
is expected to appear indirectly through the gravitational 
interaction which mediates the effects of the condensations
(\ref{cond}) \cite{gaugino}. 
In the general supergravity framework, the order parameter for the 
supersymmetry breaking in the observable sector is expressed as 
\begin{equation}
{\cal F}=
\sum_i{\cal F}^i+{\cal F}_\lambda+\cdots
=\sum_i{\partial^i W_{\rm hid}\over M_{\rm pl}}e^{K\over 2M_{\rm pl}^2}
+{1\over 4M_{\rm pl}}\langle\lambda_h\lambda_h\rangle+\cdots,
\label{fsusy}
\end{equation} 
where $K$ is the K\"ahler potential and the minimal kinetic terms are 
assumed. Since the condensations (\ref{cond}) are assumed to be 
generated in the hidden sector, the dominant supersymmetry breaking 
effect in the observable sector is expected to come from 
the first term of eq.~(\ref{fsusy}). 
In fact, the second term of $W_{\rm hid}^{\rm YM}$ generates the dominant 
effect of the condensation mediated into 
the observable sector.\footnote{The corresponding term to this
does not exist in the model of \cite{aquin} and the difference from 
our scenario is caused by this point. 
Although there may be other contributions to the
supersymmetry breaking such as the one due to the moduli $F$ term,
for example, we do not refer to it here.}
This supersymmetry breaking effect is expected to have a typical scale 
\begin{equation}
M_{\rm susy}\equiv\sqrt{\langle {\cal F}^\psi\rangle}
=O({\Lambda_h^2\over M_{\rm pl}}). 
\end{equation}
If we require $M_{\rm susy}=O(10^3)$~GeV,
$\Lambda_h$ should be an intermediate 
scale such as $\Lambda_h=O(10^{10})$~GeV. 

Since our model has the singlet chiral superfields  $\hat S_{1,2}$ which have
U(1)$_{\tilde X}$ charges, they may give additional contributions 
to the above mentioned pNG boson if their scalar components get the vacuum 
expectation values to induce the spontaneous symmetry breaking of
U(1)$_{\tilde X}$.   
We parameterize the fluctuation of the scalar components $\tilde S_{1,2}$ 
from their vacuum values as
\begin{equation}
\tilde S_1\equiv \left(u_1+\eta_1(x)\right) e^{i{\theta_1(x) \over u_1}}, 
\qquad
\tilde S_2\equiv \left(u_2+\eta_2(x)\right) e^{i{\theta_2(x) \over u_2}},
\label{svev}
\end{equation}
where $u_i=|\langle\tilde S_i\rangle|$.  Originally, there are two degrees of
freedom for the phases of $\tilde S_{1,2}$. However, one of their linear
combination $\xi\equiv (2\sqrt{3}\theta_1+3\sqrt{2}\theta_2)/5$ can be 
fixed dynamically to a constant value as a potential minimum 
during the early stage of the evolution of $\tilde S_{1,2}$.
We will discuss this point later. 
Thus, only its orthogonal state 
$\theta\equiv (-3\sqrt{2}\theta_1+2\sqrt{3}\theta_2)/5$ 
can contribute to the discussing pNG boson $\theta_{\tilde X}$. 

We make an additional assumption related to this pNG sector, which is 
natural if the model is supposed to be a low energy effective 
theory of some superstring. That is, the model is assumed to have 
a model independent axion $a_{\rm MI}$. 
Its physical property can be briefly summarized as follows.
It is a dual of the field strength
$H_{\mu\nu\rho}$ of the second rank antisymmetric tensor $B_{\mu\nu}$
and is defined by $\partial^\mu a_{\rm
MI}=\epsilon^{\mu\nu\rho\sigma}H_{\nu\rho\sigma}$ \cite{miaxion}.
Its decay constant $f_{a_{\rm MI}}$is generally considered to be 
$f_{a_{\rm MI}}=O(M_{\rm pl})$.
It constitutes an imaginary part of the dilaton and has universal effective 
couplings with non-Abelian gauge fields such as 
${1\over 32\pi^2}{a_{\rm MI}\over f_{a_{\rm MI}}}
{\cal F}^\alpha_{\mu\nu}\tilde {\cal F}^{\alpha\mu\nu}$,
where ${\cal F}^\alpha_{\mu\nu}$ is the non-Abelian gauge field strength and
$\tilde{\cal F}^\alpha_{\mu\nu}$ is its dual. 
In the present model, $a_{\rm MI}$ has this type of coupling with the
SU(3) and SU(N) gauge fields.

\section{Dark energy candidate}
In this section we show that our model can include a quintessential
axion as a dark energy candidate in the similar way to \cite{aquin}.
The model is supposed to have two axions as discussed in the previous section.
One is the pNG boson associated to the spontaneous breaking of 
U(1)$_{\tilde X}$ and another one is the model independent axion. 
Since the operators in $W_{\rm ob}$, $W_{\rm hid}$ and $W_{\rm mix}$ 
except for the first term in $W_{\rm mix}$ are U(1)$_{\tilde X}$ invariant, 
the potential for these pseudo-scalar fields is expected to be 
induced mainly from their
effective interactions with the SU(3) and SU(N) instantons.
These interactions can be written as
\begin{equation}
{2\over 32\pi^2}\left({a\over f_a}
+{a_{\rm MI}\over f_{a_{\rm MI}}}\right)
F_{\mu\nu}\tilde F^{\mu\nu}, \qquad
{2\over 32\pi^2}\left({a_h\over f_h} 
+{a_{\rm MI}\over f_{a_{\rm MI}}}\right)
G_{\mu\nu}\tilde G^{\mu\nu}, 
\label{inst}
\end{equation}
where $F_{\mu\nu}$ and $G_{\mu\nu}$ are the field strengths 
of SU(3) and SU(N) and their gauge indices are abbreviated. 
Axion fields $a$, $a_h$ and
their decay constants $f_a$, $f_h$ are defined as \cite{cp}
\begin{eqnarray}
a&=&{5\over f_a}\left(u_1\theta_1
+ n_{G_2}\Lambda_h\theta_{G_2}\right)
\simeq {5n_{G_2}\Lambda_h\over f_a}\gamma\theta_{\tilde X}+\cdots, \nonumber \\
f_a&=&5\left(u_1^2+n_{G_2}^2\Lambda_h^2
\right)^{1/2}\simeq 5n_{G_2}\Lambda_h,\nonumber\\
a_h&=&{1\over f_h}\left(5n_{G_2}\Lambda_h\theta_{G_2}
+2n_{G_1}\Lambda_h\theta_{G_1}-2\Lambda_h\theta_\lambda\right)
\simeq{5n_{G_2}\Lambda_h\over f_h\gamma}\theta_{\tilde X}, \nonumber \\
f_h&=&\left(25n_{G_2}^2\Lambda_h^2
+4n_{G_1}^2\Lambda_h^2+4\Lambda_h^2\right)^{1/2}
\simeq {5n_{G_2}\Lambda_h\over\gamma},
\end{eqnarray}
where $\gamma=\left(1+{4n_{G_1}^2+4\over 25n_{G_2}^2}\right)^{-1/2}$.
The ellipses in the expression of $a$ stand for the
components orthogonal to $\theta_{\tilde X}$. In these formulas, we assume 
$|\langle H_{1,2}\rangle|\ll u_{1,2}\ll\Lambda_h$ in the case of 
$u_{1,2}\not= 0$.
Then the contributions from the U(1)$_{\tilde X}$ breaking due to 
$\langle H_{1,2}\rangle$ and $u_{1,2}$ are neglected. 

If we assume that the SU(N) gaugino and the fermionic components 
of the chiral superfields $\hat F$ and $\hat G_{1,2}$ get 
the masses $M_{\tilde g}$, $m_{G_1}$ and $m_{G_2}$,
the scalar potential for these axions generated
through the interactions (\ref{inst}) is found to be written as \cite{aquin}
\begin{equation}
V=m_{\pi}^2f_{\pi}^2\left[1-\cos\left(\gamma{\theta_{\tilde X}\over f_h}
+{a_{\rm MI}\over M_{\rm pl}}\right)\right]
+V_h\left[1-\cos\left({\theta_{\tilde X}\over f_h}
+{a_{\rm MI}\over M_{\rm pl}}\right)\right],
\label{pot}
\end{equation}
where $V_h=m_{G_1}^{n_{G_1}}m_{G_2}^{n_{G_2}}
M_{\tilde g}^N\Lambda_h^{4-n_{G_1}-n_{G_2}-N}$ and 
we take $f_{a_{\rm MI}}=M_{\rm pl}$.
The cosmological constant is assumed to be zero.
For the observable sector, we use the ordinary invisible axion potential 
due to the QCD instanton \cite{invis}. 
One may wonder that the higher order $\tilde X$ violating 
operators ignored in $W_{\rm hid}$ and $W_{\rm mix}$ might give larger 
contribution to the $a_h$ potential compared with 
that given in eq.(\ref{pot}). 
The most dangerous operator among the $\tilde X$ breaking 
higher dimensional ones is the supersymmetry breaking operator 
of the type ${B\over M_{\rm pl}^{k-3}}(\tilde\Phi_1\cdots\tilde\Phi_k)$ 
which corresponds to the superpotential operator 
${1\over M_{\rm pl}^{k-3}}(\hat\Phi_1\cdots\hat\Phi_k)$.   
If we take $B\simeq 10^3$~GeV and $\langle\tilde\Phi\rangle
\simeq 10^{10}$~GeV as their
conservative values, these effects cannot dominate the second term of
eq.~(\ref{pot}). In fact, they determine the mass of $a_h$ 
only if $k\le 5$ is satisfied.
However, we can easily check that this is not the case in the present model. 
Thus, it is sufficient to consider
eq.~(\ref{pot}) as the dominant
potential for $a_h$.\footnote{We note that the $\tilde X$ violating but
$Z_{18}\times Z_3$ invariant higher dimensional operators which include
gauginos cannot give a dominant contribution either.}

Using the scalar potential (\ref{pot}), the mass matrix for two axions can be
written in the basis $(\theta_{\tilde X}, a_{\rm MI})$ as
\begin{equation}
\left(\begin{array}{ccc}
{1 \over f_h^2}(\gamma^2m_\pi^2f_\pi^2+V_h)&
{1 \over f_hM_{\rm pl}}(\gamma m_\pi^2f_\pi^2 +V_h)\\
{1 \over f_hM_{\rm pl}}(\gamma m_\pi^2f_\pi^2 +V_h)&
{1 \over M_{\rm pl}^2}(m_\pi^2f_\pi^2+V_h)\\
\end{array}\right).
\end{equation}
Since $\gamma\not=1$ is satisfied in the present model, we find that
both eigenvalues are nonzero and they are expressed as
\begin{equation}
m_{a_{\rm QCD}}^2\simeq {{m_\pi}^2f_\pi^2\over (f_h/\gamma)^2}, \qquad
m_{a_q}^2\simeq {m_{G_1}^{n_{G_1}}m_{G_2}^{n_{G_2}}M_{\tilde g}^N
\Lambda_h^{4-n_{G_1}-n_{G_2}-N}\over M_{\rm pl}^2}.
\label{mass}
\end{equation}
The corresponding eigenstates can be given by
\begin{equation}
a_{\rm QCD}\simeq \theta_{\tilde X}-{f_h/\gamma\over M_{\rm pl}}a_{\rm MI}, \qquad
a_q \simeq {f_h/\gamma \over M_{\rm pl}}\theta_{\tilde X} +a_{\rm MI}.
\end{equation}
Since the QCD axion $a_{\rm QCD}$ is dominated by $\theta_{\tilde X}$, 
and its decay constant is given by $f_h/\gamma$, 
it should satisfy the astrophysical constraint 
$f_h/\gamma \simeq 10^{9\sim 12}$GeV \cite{astro}.
This is found to be realized for the supposed values of $\Lambda_h$ and
$\gamma$ here. Thus, $a_{\rm QCD}$ is expected to behave as the 
invisible axion to solve the strong CP problem. 

It should be noted that both the origin of the $\mu$ term and 
the candidate of the cold dark matter are intimately related to 
the scale of $\Lambda_h$.
In the present model, we have two sources for the $\mu$ term, which are 
$\mu=\lambda_1\langle\tilde S_1\rangle$ and  $\mu ={d_2\over M_{\rm pl}} 
\langle \tilde F\tilde G_2\rangle$. 
For the first case, we should take account of the assumed 
relation $|\langle\tilde S_1\rangle|\ll \Lambda_h$. 
As far as this is satisfied, $\mu$ can take various values depending 
on the values of $\langle\tilde S_1\rangle$ and $\lambda_1$.
On the other hand, for the latter case, if we take
$\Lambda_h\simeq10^{12}$~GeV so that $a_{\rm QCD}$ can be a cold dark
matter candidate \cite{aquin}, the $\mu$ scale takes a large value such
as $O(10^6)$~GeV.
In order to make its contribution to the $\mu$ term suitable,
we should assume $\Lambda_h\sim 10^{10}$~GeV which is just a supposed
value here.
Since $f_h/\gamma$ seems to be too small for $a_{\rm QCD}$ 
to be the dark matter candidate in that
case \cite{cdmaxion}, we need another candidate for it. 
We will discuss this problem in the next section. 

The formula in eq.~(\ref{mass}) shows that the mass of $a_q$ can be 
strongly suppressed so that the present Hubble constant $H_0$
can be larger than that. Then, $a_q$ can still now stay 
near the initial value at the beginning of the universe. It 
is naturally expected that $a_q$ is far from the true vacuum.
This means that the state equation for $a_q$ can satisfy 
\begin{equation}
w={{1\over 2}\dot a_q^2- V(a_q)\over{1\over 2}\dot a_q^2 +V(a_q)}
\simeq -1 
\end{equation}
as far as $a_q$ takes large value. 
Since $a_q$ is dominated by $a_{\rm MI}$, 
it is natural to take the initial value of $a_q$ as $a_q\simeq M_{\rm pl}$.
If the energy density of $a_q$ dominates that of the present universe, 
$a_q$ controls the expansion behavior of the universe. In that case the 
observed accelerated expansion of the present universe may be explained. 
If we assume $a_q\simeq M_{\rm pl}$, the energy density stored in $a_q$ 
can be estimated as
\begin{equation}
\rho_{a_q}\simeq m_{G_1}^{n_{G_1}}m_{G_2}^{n_{G_2}}M_{\tilde g}^N
\Lambda_h^{4-n_{G_1}-n_{G_2}-N},
\label{de}
\end{equation}
where we use $m_{a_q}^2$ in eq.(\ref{mass}).

Here we assume that the SU(N) gauginos get the supersymmetry breaking mass
$M_{\tilde g}\simeq 10^3$~GeV.\footnote{In the present case we need 
to consider an additional supersymmetry breaking based on other source 
such as the moduli $F$-term breaking, 
since $\Lambda_h\simeq 10^{10}$~GeV is too small to realize the 
gaugino mass of this order through the gaugino condensation (\ref{cond}).}
The fermionic components of $\hat F$ and $\hat G_{1,2}$ can
get the masses through the couplings in $W_{\rm hid}^{\rm YM}$ and 
$W_{\rm mix}$ if we assume $\langle\tilde\phi\rangle\not= 0$ and 
$\langle\tilde\psi\rangle= 0$. 
They are expressed as
\begin{equation}
m_{G_1}=g_1{|\langle\hat\phi\rangle|^2\over M_{\rm pl}}, \qquad 
m_{G_2}=d_2{|\langle H_1\rangle\langle H_2\rangle|\over M_{\rm pl}}.
\end{equation}
The second formula results in $m_{G_2}\simeq 10^{-15}$GeV.
On the other hand, if we suppose 
$\langle\tilde\phi\rangle=O(10^{10})$~GeV, $m_{G_1} \sim 10$~GeV is 
induced.\footnote{We do not discuss the origin of nonzero
$\langle\tilde\phi\rangle$ but just assume its value in this paper.
This value of $\langle\tilde\phi\rangle$ 
is also required from another aspect of 
the model as seen in the next section.}
If we use these values in eq.(\ref{de}),
the energy density of $a_q$ can be estimated as
\begin{equation}
\rho_{a_q}\simeq 10^{40-9n_{G_1}-24n_{G_2}-7N}{\rm GeV}^4.
\end{equation}
If we use $n_{G_1}=3$, $n_{G_2}=1$ and $N=5$ which make the present model
consistent as seen in the previous section, we have 
$\rho_{a_q}\simeq 10^{-47}~{\rm GeV}^4$. This value corresponds to 
the predicted value $(0.003~{\rm eV})^4$ as the dark energy on the basis of 
the observation of the CMB anisotropy. 
This result shows that it is possible to identify $a_q$ with the
quintessential axion which is the origin of the dark energy.

Although the scenario is very similar to the one in \cite{aquin}, 
our model is extended by introducing additional chiral superfields. 
As its result, the scale of the condensations in the hidden sector can 
take a lower value than that in \cite{aquin}.
It easily makes the $\mu$ scale take an appropriate value for the electroweak 
symmetry breaking keeping the supersymmetry breaking scale in the 
observable sector to be in the desirable region. 
Moreover, it can give an interesting explanation for  
the quantitative relation between the dark matter abundance and the baryon 
number asymmetry as shown in the next section.
 
\section{Abundance of baryon and dark matter}
In this framework, both abundance of the baryon and the dark matter may be
also explained following the scenario presented in \cite{s}.
We discuss this point in this section.
We consider the early time evolution of the $D$-flat direction of the 
singlet fields $\tilde S_1$ and $\tilde S_2$ defined by 
\begin{equation}
\langle\tilde S_1\rangle={u\over\sqrt{3}}e^{i{\xi\over u}}, \qquad
\langle\tilde S_2\rangle={u\over\sqrt{2}}e^{i{\xi\over u}},
\label{vev}
\end{equation}
which is expected to be realized when they have effectively the same
negative squared mass.\footnote{Since the potential 
minimum can be realized along the subspace $3|\tilde S_1|^2=2|\tilde S_2|^2$ 
as far as the soft supersymmetry breaking masses for $\tilde S_1$ 
and $\tilde S_2$ are equal, one may expect 
$|\langle\tilde S_1\rangle|=u/\sqrt{3}$ and 
$|\langle\tilde S_2\rangle|=u/\sqrt{2}$. In the phase part of the
expression in eq.~(\ref{vev}), 
we omit the component orthogonal to $\xi$, which corresponds to 
the pNG boson associated with the U(1)$_{\tilde X}$ breaking as 
discussed in the previous section.
It is expressed as $\theta\equiv (-3\sqrt{2}\theta_1+2\sqrt{3}\theta_2)/5$. 
Its potential is induced not from $W_{\rm mix}$ but through
the SU(3) and SU(N) instanton effects.}
This direction is slightly lifted by the nonrenormalizable operators 
in the scalar potential which are induced from the first term of 
the superpotential $W_{\rm mix}$.
In the following discussion, 
the inflaton is assumed to couple similarly with the fields in the
observable and hidden sectors.
 
In the early universe, there are effective contributions 
to the scalar potential for this flat direction, which is induced 
by the supersymmetry breaking effects caused 
by the large Hubble constant $H$ \cite{dens} and the thermal effects 
\cite{temp} other than the ordinary soft supersymmetry breaking.
If we take account of these effects, the scalar potential in this 
direction is found to be expressed as 
\begin{eqnarray}
V&\simeq&\left(-cH^2+M_u^2(T)\right)u^2
+{5\vert d_1\vert^2\over 12}
{u^8\over M_{\rm pl}^4} \nonumber\\ 
&+&{1\over 6\sqrt{2}}
\left\{\left({aM_{\rm susy}e^{i\theta_a}\over M_{\rm pl}^2}
+{bHe^{i\theta_b} \over M_{\rm pl}^2}\right)
u^5e^{i{\xi\over u}}+{\rm h.c.}\right\},
\label{eqaa}
\end{eqnarray}
where $M_{\rm susy}$ is a typical soft supersymmetry breaking 
scale of $O(1)$~TeV 
and the coefficients $a$, $b$ and $c$ are $O(1)$ real constants. 
CP phases $\theta_a$ and $\theta_b$ in the curly brackets 
are induced by the above mentioned supersymmetry breaking effects 
which violate U(1)$_X$ by an amount of $\Delta X(=-4)$.  
The effective mass $M_u^2(T)$ contains the usual soft supersymmetry 
breaking mass $m_{S}^2$ of $O(M_{\rm susy}^2)$ and the thermal mass 
$C_T\lambda_{1,2}^2T^2$ caused by the 
coupling of $\hat{S}_{1,2}$ with $\hat{H}_{1,2}$ and $\hat{C}_{1,2}$ 
in the thermal plasma.\footnote{Since the effective degrees of freedom
for the relativistic fields are not same in both sectors,
the reheating temperature in each sector may be somewhat different.
However, this difference does not change the following discussion
largely.} 
It can be expressed depending on the value of $u$ as \cite{temp}
\begin{equation}
M_u^2(T)\simeq\left\{\begin{array}{ll}m^2_{S} & \quad (\lambda_{1,2} u>T), \\
m^2_{S}+C_T\lambda_{1,2}^2 T^2 & \quad (\lambda_{1,2} u<T), \\
\end{array}\right.
\label{therm}
\end{equation} 
where $C_T$ is a numerical factor for the thermal mass.

During the inflation, the Hubble constant contribution $H^2$ dominates 
the mass of the condensate in the scalar potential (\ref{eqaa}).
If the sign of this Hubble constant contribution is negative 
$(c>0) $\cite{dens}, the magnitude of the condensate takes a large 
value such as
\begin{equation}
u_I\simeq \left(H M_{\rm pl}^2\right)^{1\over 3}.
\label{eqa3}
\end{equation} 
On the other hand, the phase $\xi/u$ of the condensate takes 
one of the five distinct values 
$\xi/u=-\theta_b/5+2\pi \ell/5~(\ell=1\sim 5)$ at the potential minimum. 
Since $H$ is much larger than the mass of the condensate
at this period and then its evolution is almost the critical damping,
the condensate follows this instantaneous potential minimum.

The dilute plasma appears as a result of a partial decay of the inflaton. 
Then the temperature rapidly increases to 
$T_{\rm max}\simeq (T_R^2H M_{\rm pl})^{1/4}$ \cite{temp}.
$T_R$ is the reheating temperature realized after the completion 
of the inflaton decay and it can be expressed as 
$T_R\simeq\sqrt{M_{\rm pl}\Gamma_I}$. $\Gamma_I$ is the inflaton
decay width.
If this temperature $T_{\rm max}$ does not satisfy 
$\lambda_{1,2}\vert u_I\vert<T_{\rm max}$, 
no thermal contribution to $M_u^2(T)$ 
appears and $M_u^2(T)$ takes the expression of the upper one 
in eq.~(\ref{therm}) \cite{dens,temp}. 
Thus, we can neglect the thermal effects during the
inflation as far as the following lower bound on $\lambda_{1,2}$ is
satisfied:
\begin{equation}
\lambda_{1,2} > T_R^{1\over 2}H_I^{-1\over 12}M_{\rm pl}^{-5\over 12},
\label{eqa4}
\end{equation}
where $H_I$ is the Hubble parameter during the inflation.
In the following discussion, we assume that this is satisfied.

When $H$ decreases to $H\sim M_{\rm susy}$ as a result of the evolution of the 
inflaton,\footnote{Here we assume an inflation scenario 
such that this period is before the reheating. 
This means that $M_{\rm susy}>\Gamma_I$ 
is satisfied and then $T_R<\sqrt{M_{\rm pl}M_{\rm susy}}
\simeq 10^{11}{\rm GeV}$.}  
the effective squared mass of the condensate becomes positive and 
then $u=0$ is the minimum of the scalar potential $V$.
The condensate starts to oscillate around $u=0$, and the
thermal effects due to the dilute plasma to $M_u^2(T)$ is expected to appear. 
Then $M_u^2(T)$ takes the expression of the lower one 
in eq.~(\ref{therm}). At this time, the dominant term for the 
U(1)$_X$ breaking in eq.~(\ref{eqaa}) changes from the second term 
in the curly brackets to the first one. 
Since the phases $\theta_a$ and $\theta_b$ are generally independent,
the phase $\xi/u$ of the condensate changes non-adiabatically from 
that determined by $\theta_b$ to that determined by $\theta_a$ 
due to the torque in the angular direction. 
Thus, the $X$ asymmetry is stored in the condensate during its 
evolution due to the AD mechanism \cite{ad}.

The produced $X$ asymmetry can be estimated by taking account that 
the $X$ current conservation is violated by the dominant $X$ breaking
operator in the curly brackets in eq.~(\ref{eqaa}) as
\begin{equation}
{d\Delta n_{X}(t)\over dt}= \Delta X
{aM_{\rm susy}u^5\over M_{\rm pl}^2}\sin\delta,
\end{equation} 
where $\Delta X$ is the $X$ charge of that operator 
and $\delta$ is determined 
by the difference between $\theta_a$ and $\theta_b$. 
By solving this equation, the $X$ asymmetry produced in the condensates 
$\langle\tilde S_{1,2}\rangle$ at this period 
is found to be roughly expressed as 
\cite{ad,dens,temp}\footnote{The rigorous estimation requires the
numerical calculation as discussed in \cite{dens}. 
It is beyond the scope of this paper and we do not refer to it further here.}
\begin{equation}
\Delta n_{X}(t)\simeq \Delta X{M_{\rm susy}\over H}H^{5\over 3}
M_{\rm pl}^{4\over 3}\sin\delta,
\end{equation}
where $t$ is the time when $H\sim M_{\rm susy}$.  Following this period, the
reheating due to the inflaton decay is completed at 
$H\sim \Gamma_I$.

Since the oscillation of the condensate behaves as the matter for 
the expansion of the universe, it can dominate the energy density 
of the universe before its decay which is considered to occur 
at $H\sim \Gamma_{\tilde S_{1,2}}$.
This is the case in the present model, 
since $\Gamma_{\tilde S_{1,2}}<\Gamma_I$ is satisfied 
for the reasonable values of $\lambda_{1,2}$.
The $X$ asymmetry stored in the condensate is 
liberated into the thermal plasma in the observable sector and also into the
hidden sector through the decay of the condensate 
by the $X$ conserving couplings $\lambda_1 \hat{S}_1\hat{H}_1\hat{H}_2$ and  
$\lambda_2 \hat{S}_2\hat{C}_1\hat{C}_2$, respectively.
Although the $X$ charge is assumed to be broken through the SU(N) 
super Yang-Mills dynamics in the hidden sector,
the singlet part in the hidden sector is connected to the super
Yang-Mills part only by the gravitational interactions as shown
in $W_{\rm mix}$.  As a result, this $X$ breaking in the super
Yang-Mills part does not affect the part 
composed of the singlet chiral superfields where the $X$ charge is
exactly conserved. The condensate decays into this $X$ conserving part. 
Taking account of these features, the ratio of the $X$ asymmetry 
$\Delta n_X^{\rm ob}$ and $\Delta n_X^{\rm hid}$
liberated into each sector to the entropy density $s$ is estimated as
\begin{equation} 
Y_X^i\equiv{\Delta n_{X}^i(\tilde t_R)\over s}
={\Delta n_{X}(\tilde t_R)\over 2s}
\simeq {\Delta n_{X}(t)\over 2\tilde T_R^3}{t^2\over \tilde t_R^2}
\simeq {\Delta X\over 2}
\tilde T_RM_{\rm susy}^{-1\over 3}M_{\rm pl}^{-2\over 3}
\sin\delta, 
\label{eqe}
\end{equation}
where we use 
$\tilde t_R\sim \Gamma_{\tilde S_{1,2}}^{-1}\sim M_{\rm pl}/\tilde T_R^2$. 
We assume that the decay widths of $\tilde S_{1,2}$ satisfy
$\Gamma_{\tilde S_1}\simeq\Gamma_{\tilde S_2}$ to make the discussion 
clear. General cases will be discussed in Appendix A.

Now we discuss the physical consequence of the $X$ asymmetry
liberated in both sectors.
At first, we focus our attention to the observable sector.
If the temperature $\tilde T_R(\simeq 10^{10}\lambda_{1}{\rm GeV})$ 
is appropriate to keep the $X$ asymmetry\footnote{This 
$\tilde T_R$ is found to be a marginal value for the cosmological gravitino 
problem \cite{grav}.} and convert it into the $B$ asymmetry through 
the sphaleron interaction \cite{spha}, we can obtain the $B$ asymmetry 
 \cite{s}.
A necessary condition for this is that $\tilde T_R$ should be higher than
$T_{ss}$, at which the soft supersymmetry breaking operators start to be 
in the thermal equilibrium. If this is not satisfied, the soft
supersymmetry breaking operators wash out the $X$ asymmetry and
$\Delta n_X^{\rm ob}=0$ results in.
However, if the $X$ and $B-L$ violating interaction in the observable
sector leaves the thermal equilibrium before the temperature reaches $T_{ss}$, 
the equilibrium conditions allow to be $\Delta n_X^{\rm ob}\not= 0$ and
a part of it is converted into the $B-L$ asymmetry. 
We assume it in the following part. The detailed
discussion about the relation between $B-L$ and $\Delta n_X^{\rm ob}$ is
presented in Appendix B. 

If the $B-L$ asymmetry existing at $T_{ss}$ is kept after that,
 it can be related to the $B$ and $L$ asymmetry as in the ordinary 
MSSM case as follows:
\begin{equation}
B={4(2N_g+1)\over 22N_g+13}(B-L),
\end{equation}
where $N_g$ is the generation number of quarks and leptons.
Thus, the $B$ asymmetry produced in this scenario is finally
estimated as
\begin{equation}
Y_B\equiv {\Delta n_B\over s} \simeq {\Delta n_X\over 2s}f(N_g)\kappa 
\simeq {\Delta X\over 2}\tilde T_R~M_{\rm susy}^{-1\over 3}
M_{\rm pl}^{-2\over 3}f(N_g)~\kappa\sin\delta,
\label{eqe4}
\end{equation}
where eq.~(\ref{eqe}) is used and $\kappa (\le 1)$ is introduced 
to take account of the washout effect.
$f(N_g)$ is a numerical factor defined by
\begin{equation}
f(N_g)={B-L\over X}~{4(2N_g+1)\over 22N_g+13},
\end{equation}
and it takes $f(3)\simeq 0.3$ for $N_g=3$.
Using this fact in eq.~(\ref{eqe4}), we find that this scenario can 
produce the presently observed $B$ asymmetry $Y_B=(0.6\sim 1)\times 10^{-10}$ 
as far as $\tilde T_R~{^>_\sim}~10^4/(\Delta X\kappa\sin\delta)$~GeV 
is satisfied.

It is useful to comment on the relation of this scenario 
to the ordinary thermal
leptogenesis here. In the seesaw model, the leptogenesis is usually 
considered on the basis of the out-of-equilibrium decay of the 
heavy right-handed 
neutrinos \cite{lept} or the decay of sneutrino condensate \cite{sneu}. 
However, if we consider the $X$ charge asymmetry generation 
along the almost flat direction of $\langle\tilde S_1\rangle$ 
as discussed above, 
the $B$ asymmetry produced by this usual leptogenesis 
might not be the dominant one. As mentioned before, we assume that 
the decay of the condensate is completed at the temperature above
$T_{ss}$, which can be sufficiently lower
than the masses of the right-handed heavy neutrinos.
Then the $B$ asymmetry produced through the usual 
scenario seems to be washed out or overridden by the $B$ asymmetry 
produced in the present scenario.\footnote{Even in
the ordinary scenario based on the heavy right-handed (s)neutrino decay,
it may be possible to make the reheating temperature low enough such as
$T_R\simeq 10^6$~GeV if the neutrino mass texture satisfies
suitable features \cite{bs}. In such a case both sources of the $B$
asymmetry may compete.}

Next we consider the hidden sector.
Our interest is whether the $X$ asymmetry liberated into 
the hidden sector through the coupling $\lambda_2\hat S_2\hat C_1\hat C_2$ 
can explain the dark matter abundance.
Since there is no renormalizable U(1)$_X$ violating interaction in the
singlet part and also no renormalizable interaction converts 
the $X$ asymmetry in the singlet part into the SU(N) super Yang-Mills part
or the observable sector, the $X$ asymmetry is considered to be conserved 
exactly within the part composed of the singlet chiral superfields
as far as no scalar components with the $X$ charge 
get the vacuum expectation values.
These features suggest that the lightest fermionic components of the
singlet part can be the dark matter candidate.
Since $\hat\phi$ has zero $X$ charge, the $X$ charge conservation is not
broken even in the case of $\langle\tilde\phi\rangle\not=0$.
Thus, if $\langle\tilde\phi\rangle\not=0$ is realized,
fermionic components of the singlet fields can get mass through the 
interactions in $W_{\rm hid}^S$ without violating the $X$ charge
conservation. Mass terms of these fermions can be written as
\begin{eqnarray}
&&(C_1~ \bar C_3 )\left(\begin{array}{cc}
0 & f_1{\langle\tilde\phi\rangle^2\over M_{\rm pl}}\\
f_1{\langle\tilde\phi\rangle^2\over M_{\rm pl}} & 0 \\\end{array}
\right)\left(\begin{array}{c}\bar C_1 \\ C_3 \\\end{array}\right)
+(C_2~ \bar C_4 )\left(\begin{array}{cc}
0 & f_2{\langle\tilde\phi\rangle^2\over M_{\rm pl}}\\
f_2{\langle\tilde\phi\rangle^2\over M_{\rm pl}} & 0 \\\end{array}
\right)\left(\begin{array}{c}\bar C_2 \\ C_4 \\\end{array}\right) \nonumber\\
&+&(S_2~ \bar C_5 )\left(\begin{array}{cc}
0 & f_3{\langle\tilde\phi\rangle^2\over M_{\rm pl}}\\
f_3{\langle\tilde\phi\rangle^2\over M_{\rm pl}} & 0 \\\end{array}
\right)\left(\begin{array}{c}\bar S_2 \\ C_5 \\\end{array}\right).
\end{eqnarray}
This shows that every fermionic components relevant to the
singlet part $W_{\rm hid}^S$ have masses 
$m_{\rm LP}=O({\langle \tilde\phi\rangle^2\over M_{\rm pl}})$ and 
some of them constitute the lightest stable one to which the $X$ 
charge asymmetry are finally distributed. 

Taking account of these aspects and eqs.~(\ref{eqe}) and (\ref{eqe4}), 
we can estimate the ratio of the baryon energy density to this 
dark matter energy density in such a way as
\begin{equation}
{\Omega_B\over \Omega_{\rm DM}}\simeq
{m_pY_B\over m_{\rm LP} Y_X^{\rm hid}}\simeq f(3)\kappa{m_p\over m_{\rm LP}},
\end{equation} 
where $\Omega_i$ is the ratio of the energy density $\rho_i$ to the
critical energy density $\rho_{\rm cr}$ in the present universe.
Mass of the proton is represented by $m_p$.
This relation suggests that the presently observed 
value $\Omega_B/\Omega_{\rm CDM}\sim 0.17$ can be explained 
if $m_{\rm LP}$ is the similar order value to $m_p$ 
in the case of $\kappa=O(1)$. 
This value of $m_{\rm LP}$ can be realized for 
$\langle\tilde\phi\rangle=O(10^{9-10})$~GeV.  
It is interesting to note that this $\langle\tilde\phi\rangle$ is the
same order as the predicted one for the explanation of the dark energy 
in the previous section. In the present scenario the dark matter mass is
related not only to $\Omega_B/\Omega_{\rm DM}$ but also to the amount of
the dark energy. This may give a viewpoint to approach the tuning problem of
the dark matter mass, which is mentioned in the footnote in the 
introduction.

Finally, we give an additional comment on the $\mu$ term.
As mentioned in section 3, the present model has two sources 
for the $\mu$ term.
Here we note that the late time evolution of $\langle\tilde S_1\rangle$ may 
give important contribution to the $\mu$ term \cite{s}.
We assume $H_I\sim 10^{13}$~GeV during the inflation on the basis of the CMB 
data and $T_R~{^<_\sim}~10^9$~GeV. For these values we obtain 
$T_{\rm max}\sim 10^{13}$~GeV. Thus, eq.~(\ref{eqa3}) 
gives $u_I\sim 10^{16}$~GeV and eq.~(\ref{eqa4}) suggests 
that $\lambda_1~{^>_\sim}~ 10^{-9}T_R^{1/2}$ should be satisfied. 
When the temperature decreases from $T_{\rm max}$ to 
$T_c\sim m_{3/2}/\lambda_1$, $M_u^2(T)$ represented by the
lower one in eq.~(\ref{therm}) starts to be dominated by the
 soft supersymmetry breaking mass $m_{\tilde S_1}^2$.
If $m_{\tilde S_1}^2<0$ is realized by some reason \cite{rad,rad1},
$\langle\tilde S_1\rangle_0\not= 0$ becomes the true vacuum after this period. 
Since the $\mu$ term is generated from the last term in 
$W_{\rm ob}$ as $\mu=\lambda_1 \langle\tilde S_1\rangle_0$,  
$\langle\tilde S_1\rangle_0$ should be in the range 
$|\langle\tilde S_1\rangle_0|~{^<_\sim}~10^{11}T_R^{-1/2}$~GeV to realize the
appropriate $\mu$ scale for the above mentioned $\lambda_1$.\footnote{Such a
value of $\langle\tilde S_1\rangle_0$ may be expected to be determined 
either by the nonrenormalizable terms or by the pure radiative 
symmetry breaking effect as discussed \cite{rad1}.}
Although the condensate again starts to oscillate around 
$\langle\tilde S_1\rangle_0$, it instantaneously decays into the light fields 
through the $X$ conserving coupling $\hat{S}_1\hat{H}_1\hat{H}_2$ 
since $H<\Gamma_{\tilde S_1}$ is satisfied at this time.  
The released energy cannot dominate the total energy density
$({\pi^2\over 30}g_\ast T^4 \gg m_{3/2}^2|\langle\tilde S_1\rangle_0|^2)$ and 
then the effects of the produced entropy is negligible.
Thus, even in this case the $X$ asymmetry obtained in eq.~(\ref{eqe}) 
can be used as the origin of the $B$ asymmetry. 

\section{Summary}
We have studied a possibility that all of the baryon number asymmetry, 
the dark matter and dark energy abundance might be related to the
extended global Abelian symmetries in a supersymmetric model. 
The introduction of some singlet chiral superfields into the MSSM
enlarges the global symmetry of the system. They may be taken as
the Peccei-Quinn type symmetry U(1)$_{\rm PQ}$ and the $R$-symmetry U(1)$_R$ 
or their linear combinations U(1)$_{\tilde X}$ and U(1)$_X$. 
If these global symmetries are shared in the observable and hidden
sectors, they may present the unified explanation for these quantities.
In this paper, such a kind of example has been concretely constructed. 
By using that model, we have shown that these symmetries could be 
a common source for them. 
In this scenario, the hidden sector is required to have two parts, which
are connected only by the gravitational interaction. 
One of them is required to break 
the global symmetry U(1)$_{\tilde X}$ explicitly through the SU(N) 
instanton effect for the dark energy
explanation. Another part to prepare the massive dark matter
should conserve U(1)$_X$ exactly.  
The flat directions associated with the introduced singlet 
fields play the essential role to produce the $X$ charge asymmetry.

The dark energy is explained as the quintessential 
axion. It is considered to appear in the model as 
the mixture state of the pNG boson 
associating with the spontaneous symmetry breaking of the 
global symmetry U(1)$_{\tilde X}$ and the model 
independent axion given by the superstring.  
The intermediate scale introduced relating to the U(1)$_{\tilde X}$ breaking 
can successfully generate the dark energy scale. It is also 
suitable for the explanation of the strong CP problem due to the 
invisible axion. In our case this invisible axion cannot be a dark
matter candidate. The operators which play the important role for this
explanation also produce the $\mu$ term with the suitable scale.

Both the baryon and the dark matter are considered to be produced 
from the charge asymmetry of U(1)$_X$. 
This asymmetry is stored in the condensate of the singlet scalar
components due to the AD mechanism. A part of this asymmetry 
is converted into the $B$ asymmetry in the observable sector.
Another part is liberated into the
hidden sector and causes the dark matter abundance. 
We have shown that both observed values of $Y_B$ and 
$\Omega_B/\Omega_{\rm CDM}$ can be explained by them in a consistent
way.

Since the dark matter in this model interacts with the fields in the
observable sector only through the gravity, it can be distinguished 
from the ordinary dark matter candidates such as the neutralino and the
axion.  However, its detection is difficult. The experimental signature
of the model might be found in the phenomena related to the coupling
$\lambda_1\hat S_1\hat H_1\hat H_2$, that is, the phenomena including
the neutral Higgs scalar or the neutralinos. To proceed such a study, it
is necessary to fix the model in a more detailed way.

In this paper we assume the rather simple structure for the hidden
sector, in particular, for the part required for the dark matter 
mass generation.  
In the present framework, however, it may be possible and also 
natural to suppose 
the similar structure to the observable sector for the 
hidden sector except for the SU(N) super Yang-Mills part.
It may constitute the mirror world for the observable sector. 
In that case the dark matter abundance may be explained just in the same
way as the $B$ asymmetry in this scenario.
The study of this kind of possibility may be worth to be proceeded.

\vspace*{5mm}
\noindent
I would like to thank the organizers of SI2005 which was held 
at Fuji-Yoshida in Japan,
during which a part of this work has been done. 
This work is supported in part by a Grant-in-Aid for Scientific 
Research (C) from Japan Society for Promotion of Science
(No.~17540246).

\newpage
\noindent
{\Large\bf Appendix A}

\noindent
In this appendix we discuss the amount of $X$ asymmetry liberated into
each sector for the general cases with arbitrary $\Gamma_{\tilde S_1}$ and 
$\Gamma_{\tilde S_2}$.
If we take account that the decay products of the condensate behave as
the radiation and also no photon is produced through its decay into the 
hidden sector, we can estimate the $X$ asymmetry 
$\Delta n_X^{\rm ob}(\tilde t_R)$ generated in the observable sector in case
$\Gamma_{\tilde S_1}>\Gamma_{\tilde S_2}$ as,
\begin{eqnarray} 
{\Delta n_{X}^{\rm ob}(\tilde t_R)\over s}&=&
{\Delta n_{X}(t)\over s}
{\Gamma_{\tilde S_1}\over \Gamma_{\tilde S_1}+\Gamma_{\tilde S_2}}
\left({t\over t_1}\right)^2,
\end{eqnarray}
and also in case $\Gamma_{\tilde S_1}<\Gamma_{\tilde S_2}$ as,
\begin{eqnarray} 
{\Delta n_{X}^{\rm ob}(\tilde t_R)\over s}&=&
{\Delta n_{X}(t)\over s}
{\Gamma_{\tilde S_1}\over \Gamma_{\tilde S_1}+\Gamma_{\tilde S_2}}
\left({t\over t_2}\right)^2\left({t_2\over t_1}\right)^{3/2},
\end{eqnarray}
where $\tilde t_R\simeq t_1=\Gamma_{\tilde S_1}^{-1}$ 
and $t_2= \Gamma_{\tilde S_2}^{-1}$.
Although  $Y_B$ has the similar expression as that discussed in the text
for the case of $\Gamma_{\tilde S_1}>\Gamma_{\tilde S_2}$, 
there appears an additional suppression factor 
$(\Gamma_{\tilde S_1}/\Gamma_{\tilde S_2})^{1/2}$ for the case of 
$\Gamma_{\tilde S_1}<\Gamma_{\tilde S_2}$,
\begin{equation}
Y_B\simeq \Delta X~ \tilde T_R~M_{\rm susy}^{-1\over 3}
M_{\rm pl}^{-2\over 3}\left({\Gamma_{\tilde S_1}\over
\Gamma_{\tilde S_2}}\right)^{1/2}
f(N_g)~\kappa\sin\delta.
\end{equation} 
The higher reheating temperature $\tilde T_R$ is required 
to produce the same $Y_B$ in comparison with the case of 
$\Gamma_{\tilde S_1}~{^>_\sim}~\Gamma_{\tilde S_2}$.

The ratio of the energy density of the baryon and the
dark matter can be expressed in both cases as
\begin{equation}
{\Omega_B\over \Omega_{\rm DM}}\simeq
{m_pY_B\over m_{\rm LP}Y_{X}^{\rm hid}}\simeq f(3)\kappa
{m_p\over m_{\rm LP}}{\Gamma_{\tilde S_1}\over\Gamma_{\tilde S_2}}.
\end{equation} 
This suggests that the presently observed value of
$\Omega_B/\Omega_{\rm DM}$ seems to be explained as far as 
$m_{\rm LP}\simeq (\Gamma_{\tilde S_1}/\Gamma_{\tilde S_2})\kappa$~GeV 
is satisfied. 
Thus, in the present model $\Gamma_{\tilde S_1}>\Gamma_{\tilde S_2}$ 
is favored from the view point of the explanation of the dark energy.

\vspace*{1cm}
\noindent
{\Large\bf Appendix B}\\
In this Appendix we discuss how the $X$ asymmetry in the observable 
sector is connected to the $B$ asymmetry \cite{s}. 
For this study, we use the detailed balance equations of relevant 
interactions \cite{lb,bp}. The particle-antiparticle
number asymmetry $\Delta n_f$ in the case of $\mu_f \ll T$ 
can be approximately represented as
\begin{equation}
\Delta n_f\equiv n_{f} -n_{f^c}=\left\{ \begin{array}{ll}
\displaystyle
{g_f\over 6}T^2\mu_f & (f~:~{\rm fermion}),\\
\displaystyle
{g_f\over 3}T^2\mu_f & (f~:~{\rm boson}), \\ \end{array} \right. 
\label{eqee}
\end{equation}
where $\mu_f$ is chemical potential and $g_f$ is a number of 
relevant internal degrees of freedom of the field $f$.
By solving the detailed valance equations for the chemical potential $\mu_f$,
we can estimate the charge asymmetry.

If the SU(2) and SU(3) sphaleron interactions are in the thermal
equilibrium, we have the conditions such as
\begin{equation}
\sum_{i=1}^{N_g}\left(3\mu_{Q_i}+\mu_{L_i}\right)+\mu_{\tilde{H_1}}
+\mu_{\tilde{H_2}}+ 4\mu_{\tilde W}=0,
\quad  
\sum_{i=1}^{N_g}\left(2\mu_{Q_i}-\mu_{U_i}-\mu_{D_i}\right)
+6\mu_{\tilde g}=0,
\label{spha}
\end{equation}
where $N_g$ is a number of the generation of quarks and leptons.
The cancellation of the total hypercharge or the electric charge 
of plasma in the universe requires
\begin{eqnarray}
&&\sum_{i=1}^{N_g}\left(\mu_{Q_i}+2\mu_{U_i}-\mu_{D_i}-\mu_{L_i}-
\mu_{E_i}\right)+\mu_{\tilde{H_2}}-\mu_{\tilde{H_1}} \nonumber \\
&&\hspace{2cm}+2\sum_{i=1}^N(\mu_{\tilde{Q_i}}+2\mu_{\tilde{U_i}}-
\mu_{\tilde{D_i}}-\mu_{\tilde{L_i}}-\mu_{\tilde{E_i}})
+2\left(\mu_{H_2}-\mu_{H_1} \right)=0.
\end{eqnarray}
When Yukawa interactions in $W_{\rm ob}$ are in the thermal 
equilibrium, they impose the following conditions:\footnote{
We should note that each term in $W_{\rm mix}$ leaves the thermal equilibrium 
at $T\sim M_{\rm pl}$. Since $\hat{S}_1$ has no other coupling to the MSSM
contents than $\lambda_1 \hat{S}_1\hat{H}_1\hat{H}_2$, 
the last one in eq.~(\ref{chemi}) is
the only condition for $\mu_{S_1}$.}
\begin{eqnarray} 
&&\mu_{Q_i}-\mu_{U_j}+\mu_{H_2}=0, \qquad
\mu_{Q_i}-\mu_{D_j}+\mu_{H_1}=0, \nonumber \\
&&\mu_{L_i}-\mu_{E_j}+\mu_{H_1}=0, \qquad
\mu_{S_1} +\mu_{\tilde H_1}+\mu_{\tilde H_2}=0.
\label{chemi}
\end{eqnarray}
There are also the conditions for the gauge interactions in the
thermal equilibrium, which are summarized as
\begin{equation}
\mu_{\tilde{Q_i}}=\mu_{\tilde g}+\mu_{Q_i}=\mu_{\tilde W}+\mu_{Q_i}
=\mu_{\tilde B}+\mu_{Q_i},
\label{eqee2}
\end{equation}
where $\mu_{\tilde g}$, $\mu_{\tilde W}$ and $\mu_{\tilde B}$ stand for
gauginos in the MSSM.
The similar relations to eq.~(\ref{eqee2}) is satisfied 
for leptons $\hat{L}_i$, doublet Higgs fields $\hat{H}_{1,2}$ and other
fields $\hat{U}_i, \hat{D}_i,  \hat{E}_i $ which have the SM 
gauge interactions.
Flavor mixings of quarks and leptons due to the Yukawa couplings
allow us to consider the flavor independent chemical potential such as
$\mu_{Q}=\mu_{Q_i}$ and $\mu_{L}=\mu_{L_i}$.

If an operator violating both $B-L$ and $X$ exists, only a linear 
combination of these two U(1)s is 
absolutely conserved. Then a part of $X$ asymmetry can be converted
into the $B-L$ asymmetry.
We consider an effective operator $(\hat{L}\hat{H}_2)^2$ for such an example.
\footnote{It corresponds to the effective neutrino mass operator 
induced in the ordinary seesaw mechanism. 
It is obtained from the operator $\hat{N}\hat{L}\hat{H}_2$ 
by integrating out the heavy right-handed neutrino $\hat{N}$.}
The thermal equilibrium condition of this operator can be written as 
\begin{equation}
\mu_L+\mu_{H_2}=0.
\label{vlep}
\end{equation}
We have to assume that this effective operator
$(\hat{L}\hat{H}_2)^2$ leaves the thermal equilibrium before the time 
when the soft supersymmetry breaking operators start to be in the thermal 
equilibrium as discussed in the text. 
Since the rate for the soft supersymmetry breaking
operators is estimated as $\Gamma_{ss}\simeq m_{3/2}^2/T$ \cite{iq},
this requires that $H>T^3/M_R^2$ is satisfied for the temperature 
obtained from the condition $H~{^>_\sim}~\Gamma_{ss}$. $M_R$ stands for the
right-handed neutrino mass. 
Since such a temperature should be 
$T~{^>_\sim}~T_{ss}\simeq 10^7$~GeV,\footnote{For 
$T~{^<_\sim}~T_{ss}$, $\mu_{\tilde g}=0$ is satisfied and then 
eqs.~(\ref{spha}) $\sim$ (\ref{vlep}) result in $\mu_{\tilde H_2}=0$. 
The $X$ asymmetry produced in the observable sector through the decay of
the condensate disappears.}, the condition is found to require
$M_R$ to satisfy $M_R~{^>_\sim}~10^{12}$~GeV.  This value of $M_R$ is 
suitable for the explanation of the light neutrino masses 
required by the neutrino oscillation data \cite{sol,atm}.
If this condition is satisfied, we can have an independent chemical 
potential from the thermal equilibrium conditions 
(\ref{spha})$\sim$(\ref{vlep}). It can be taken as $\mu_{\tilde H_2}$, 
which can be related to the above mentioned remaining symmetry.

By solving eqs.~(\ref{spha}) $\sim$ (\ref{vlep}),
$\mu_Q$, $\mu_L$, $\mu_{H_{1,2}}$ and $\mu_{\tilde g}$ can be written 
with the chemical potential of Higgsino field $\tilde H_2$ 
at the temperature $T(~{^>_\sim}~T_{ss})$ in such a way as
\begin{eqnarray}
&&\mu_Q={17N_g+6\over N_g(10N^2_g-17N_g-15)}\mu_{\tilde H_2}, \quad
\mu_L=-\mu_{H_2}={5(4N_g+3)\over 10N^2_g-17N_g-15}\mu_{\tilde H_2}, 
\nonumber \\
&&\mu_{H_1}=-{40N_g+3\over 10N^2_g-17N_g-15}\mu_{\tilde H_2}, \quad
\mu_{\tilde g}=-{(10N_g+3)N_g\over 10N^2_g-17N_g-15}\mu_{\tilde H_2}.
\label{eqee3}
\end{eqnarray}
By defining $B$ and $L$ as $\Delta n_B\equiv BT^2/6$ 
and $\Delta n_L\equiv LT^2/6$, 
we can calculate these values at $T_{ss}$ by using eqs.~(\ref{eqee})
and (\ref{eqee3}). Using the results, we can relate the $X$ asymmetry in the
observable sector to the $B-L$ asymmetry as follows:
\begin{equation}
B-L={20N_g^3+162N_g^2-24N_g-72\over 360N_g^3
+3308N^2_g-1419N_g-1143}X, 
\label{eqeed}
\end{equation}
where $X$ stands for the $X$ asymmetry liberated into the observable
sector which is defined by $\Delta n^{\rm ob}_X\equiv XT^2/6$.
After the freeze-out of $(LH_2)^2$ at the temperature
$T({^>_\sim}~T_{ss})$, the $B-L$ asymmetry is kept to be conserved.

\newpage

\end{document}